# Comment: Understanding OR, PS and DR

**Zhiqiang Tan**

We congratulate Kang and Schafer (KS) on their excellent article comparing various estimators of a population mean in the presence of missing data, and thank the Editor for organizing the discussion. In this communication, we systematically examine the propensity score (PS) and the outcome regression (OR) approaches and doubly robust (DR) estimation, which are all discussed by KS. The aim is to clarify and better our understanding of the three interrelated subjects.

Sections 1 and 2 contain the following main points, respectively.

(a) OR and PS are two approaches with different characteristics, and one does not necessarily dominate the other. The OR approach suffers the problem of *implicitly* making extrapolation. The PS-weighting approach tends to yield large weights, *explicitly* indicating uncertainty in the estimate.

(b) It seems more constructive to view DR estimation in the PS approach by incorporating an OR model rather than in the OR approach by incorporating a PS model. Tan's (2006) DR estimator can be used to improve upon any initial PS-weighting estimator with both variance and bias reduction.

Finally, Section 3 presents miscellaneous comments.

## 1. UNDERSTANDING OR AND PS

For a population, let $X$ be a vector of (pretreatment) covariates, $T$ be the treatment status, $Y$ be the observed outcome given by $(1-T)Y_0 + TY_1$, where $(Y_0, Y_1)$ are potential outcomes. The observed data consist of independent and identically distributed copies $(X_i, T_i, Y_i)$, $i = 1, \ldots, n$. Assume that $T$ and $(Y_0, Y_1)$ are conditionally independent given $X$. The objective is to estimate

$$\mu_1 = E(Y_1),$$
$$\mu_0 = E(Y_0),$$

and their difference, $\mu_1 - \mu_0$, which gives the average causal effect (ACE). KS throughout focused on the problem of estimating $\mu_1$ from the data $(X_i, T_i, T_i Y_i)$, $i = 1, \ldots, n$, only, noting in Section 1.2 that estimation of the ACE can be separated into independent estimation of the means $\mu_1$ and $\mu_0$. We shall in Section 3 discuss subtle differences between causal inference and solving two separate missing-data problems, but until then we shall restrict our attention to estimation of $\mu_1$ from $(X_i, T_i, T_i Y_i)$ only.

*Zhiqiang Tan is Assistant Professor, Department of Biostatistics, Bloomberg School of Public Health, Johns Hopkins University, 615 North Wolfe Street, Baltimore, Maryland 21205, USA e-mail: ztan@jhsph.edu.*

The model described at this stage is completely nonparametric. No parametric modeling assumption is made on either the regression function $m_1(X) = E(Y|T=1, X)$ or the propensity score $\pi(X) = P(T=1|X)$. Robins and Rotnitzky (1995) and Hahn (1998) established the following fundamental result for semiparametric (or more precisely, nonparametric) estimation of $\mu_1$.

PROPOSITION 1. *Under certain regularity conditions, there exists a unique influence function, which hence must be the efficient influence function, given by*

$$\tau_1 = \frac{T}{\pi(X)} Y - \mu_1 - \left(\frac{T}{\pi(X)} - 1\right) m_1(X)$$
$$= m_1(X) - \mu_1 + \frac{T}{\pi(X)}(Y - m_1(X)).$$

*The semiparametric variance bound (i.e., the lowest asymptotic variance any regular estimator of $\mu_1$ can achieve) is $n^{-1} E^2(\tau_1)$.*

The semiparametric variance bound depends on both $m_1(X)$ and $\pi(X)$. The bound becomes large or even infinite *whenever* $\pi(X) \approx 0$ for some values of $X$. Intuitively, it becomes difficult to infer the overall mean of $Y_1$ in this case, because very few values of $Y_1$ are observed among subjects with $\pi(X) \approx 0$. The difficulty holds *whatever* parametric approach, OR or PS, is taken for inference, although the symptoms can be different. This point is central to our subsequent discussion.

The problem of estimating $\mu_1$ is typically handled by introducing parametric modeling assumptions on either $m_1(X)$ or $\pi(X)$. The OR approach is to specify an OR model, say $m_1(X; \alpha)$, for $m_1(X)$ and then estimate $\mu_1$ by

$$\hat{\mu}_{OR} = \frac{1}{n} \sum_{i=1}^{n} \hat{m}_1(X_i),$$

where $\hat{m}_1(X)$ is the fitted response. The PS approach is to specify a PS model, say $\pi(X; \gamma)$, for $\pi(X)$ and then estimate $\mu_1$ by

$$\hat{\mu}_{IPW} = \frac{1}{n} \sum_{i=1}^{n} \frac{T_i Y_i}{\hat{\pi}(X_i)}$$

or

$$\sum_{i=1}^{n} \frac{T_i Y_i}{\hat{\pi}(X_i)} \Big/ \sum_{i=1}^{n} \frac{T_i}{\hat{\pi}(X_i)},$$

where $\hat{\pi}(X)$ is the fitted propensity score. The idea of inverse probability weighting (IPW) is to recover the joint distribution of $(X, Y_1)$ by attaching weight $\propto \hat{\pi}^{-1}(X_i)$ to each point in $\{(X_i, Y_i) : T_i = 1\}$ (see Tan, 2006, for a likelihood formulation). More generally, consider the following class of augmented IPW estimators $\hat{\mu}_{AIPW} = \hat{\mu}_{AIPW}(h)$ depending on a *known* function $h(X)$:

$$\hat{\mu}_{AIPW} = \frac{1}{n} \sum_{i=1}^{n} \frac{T_i Y_i}{\hat{\pi}(X_i)} - \frac{1}{n} \sum_{i=1}^{n} \left( \frac{T_i}{\hat{\pi}(X_i)} - 1 \right) h(X_i).$$

A theoretical comparison of the two approaches is given by

PROPOSITION 2. *Assume that an OR model is correctly specified and $m_1(X)$ is efficiently estimated with adaptation to heteroscedastic* $\text{var}(Y_1|X)$, *and that a PS model is correctly specified and $\pi(X)$ may or may not be efficiently estimated. Then*

$$\text{asy.var} (\hat{\mu}_{OR}) \leq \text{asy.var} (\hat{\mu}_{AIPW}),$$

*where* asy.var. *denotes asymptotic variance as* $n \to \infty$.

In fact, the asymptotic variance of $\hat{\mu}_{OR}$, which is the lowest under the parametric OR model, is no greater than the semiparametric variance bound under the nonparametric model, whereas that of $\hat{\mu}_{AIPW}$ is no smaller than $n^{-1} E^2(\tau_1)$ because $\tau_1$ has the smallest variance among $\pi^{-1}(X)TY - (\pi^{-1}(X)T - 1)h(X)$ over all functions $h(X)$. In the degenerate case where $m_1(X)$ and $\pi(X)$ are known, the comparison can be attributed to Rao–Blackwellization because $E[\pi^{-1}(X)TY - (\pi^{-1}(X)T - 1)h(X)|X] = m_1(X)$. This result has interesting implications for understanding the two approaches.

First, the result formalizes the often-heard statement that the (A)IPW estimator is no more efficient than the OR estimator. If a correct OR model and a correct PS model were placed in two black boxes, respectively, and if a statistician were asked to open one and only one box, then the statistician should choose the box for the OR model in terms of asymptotic efficiency (minus the complication due to adaptation to heteroscedastic variance of $Y_1$ given $X$). However, one could immediately argue that this comparison is only of phantom significance, because all models (by human efforts) are wrong (in the presence of high-dimensional $X$) and therefore the hypothetical situation never occurs. In this sense, we emphasize that the result does *not* establish any absolute superiority of the OR approach over the PS approach.

Second, even though not implying one approach is better than the other, the result does shed light on different characteristics of the two approaches as an approximation to the ideal nonparametric estimation. Typically, increasingly complicated but nested parametric models can be specified in either approach to reduce the dependency on modeling assumptions. For a sequence of OR models, the asymptotic variance of $\hat{\mu}_{OR}$ is increasing to the semiparametric variance bound, whereas for a sequence of PS models, the asymptotic variance of $\hat{\mu}_{AIPW}$ is decreasing to the semiparametric variance bound. For this difference, we suggest that the OR approach is aggressive and the PS approach is conservative. Correctly specifying an OR model ensures that $\hat{\mu}_{OR}$ is consistent and has asymptotic variance no greater, whereas correctly specifying a PS model ensures that $\hat{\mu}_{AIPW}$ is consistent and has asymptotic variance no smaller, than otherwise would be best attained without any modeling assumption. This interpretation agrees with the finding of Tan (2006) that the OR approach works directly with the usual likelihood, whereas the PS approach retains part of all information and therefore ignores other part on the joint distributions of covariates and potential outcomes.

Now the real, hard questions facing a statistician are:

(a) Which task is more likely to be accomplished, to correctly specify an OR model or a PS model?

(b) Which mistake (even a mild one) can lead to worse estimates, misspecification of an OR model or a PS model?

First of all, it seems that no definite comparison is possible, because answers to both questions depend on unmeasurable factors such as the statistician's effort and experience for question (a) and the degree and direction of model misspecification for question (b). Nevertheless, some informal comparisons are worth considering.

Regarding question (a), a first answer might be "equally likely," because both models involve the same vector of explanatory variables $X$. However, the two tasks have different forms of difficulties. The OR-model building works on the "truncated" data $\{(X_i, Y_i) : T_i = 1\}$ within treated subjects. Therefore, any OR model relies on extrapolation to predict $m_1(X)$ at values of $X$ that are different from those for most treated subjects [i.e., $\pi(X) \approx 0$]. The usual model checking is not capable of detecting OR-model misspecification, whether mild or gross, in this region of $X$. (Note that finding high-leverage observations can point to the existence of such a region of $X$, *not* model misspecification.) This problem holds for low- or high-dimensional $X$, and is separate from the difficulty to capture $m_1(X)$ within treated subjects when $X$ is high-dimensional [cf. KS's discussion below display (2)]. In contrast, the PS-model building works on the "full" data $\{(X_i, T_i)\}$ and does not suffer the presence of data truncation, although suffering the same curse of dimensionality. The exercise of model checking is capable of detecting PS-model misspecification. The matter of concern is that successful implementation is difficult when $X$ is high-dimensional.

Regarding question (b), KS (Section 2.1) suggested that the (A)IPW estimator is sensitive to misspecification of the PS model when $\pi(X) \approx 0$ for some values of $X$. For example, if $\pi(X) = 0.01$ is underestimated at 0.001, then, even though the absolute bias is small ($= 0.009$), the weight $\pi^{-1}(X)$ is overestimated by 10 times higher. In this case, the estimator has inflated standard error, which can be much greater than its bias. In contrast, if the OR model is misspecified, then the bias of the OR estimator is the average of those of $\hat{m}_1(X)$ across individual subjects in the original scale, and can be of similar magnitude to its standard deviation.

In summary, OR and PS are two approaches with different characteristics. If an OR model is correctly specified, then the OR estimator is consistent and has asymptotic variance no greater than the semiparametric variance bound. Because of data truncation, any OR model suffers the problem of *implicitly* making extrapolation at values of $X$ with $\pi(X) \approx 0$. Finding high-leverage observations in model checking can point to the existence of such values of $X$. In contrast, the PS approach specifically examines $\pi(X)$ and addresses data truncation by weighting to recover the joint distribution of $(X, Y_1)$. The weights are necessarily large for treated subjects with $\pi(X) \approx 0$, in which case the standard error is large, *explicitly* indicating uncertainty in the estimate. If a PS model is correctly specified, then the (A)IPW estimator is consistent and has asymptotic variance no smaller than the semiparametric variance bound.

## 2. UNDERSTANDING DR

The OR or the (A)IPW estimator requires specification of an OR or a PS model, respectively. In contrast, a DR estimator uses the two models in a manner such that it remains consistent if either the OR or the PS model is correctly specified. The prototypical DR estimator of Robins, Rotnitzky and Zhao (1994) is

$$\begin{aligned}
\hat{\mu}_{AIPW,\text{fix}} &= \frac{1}{n} \sum_{i=1}^{n} \frac{T_i Y_i}{\hat{\pi}(X_i)} \\
&\quad - \frac{1}{n} \sum_{i=1}^{n} \left( \frac{T_i}{\hat{\pi}(X_i)} - 1 \right) \hat{m}_1(X_i) \\
&= \frac{1}{n} \sum_{i=1}^{n} \hat{m}_1(X_i) \\
&\quad + \frac{1}{n} \sum_{i=1}^{n} \frac{T_i}{\hat{\pi}(X_i)} (Y - \hat{m}_1(X_i)).
\end{aligned}$$

The two equivalent expressions [resp. (9) and (8) in KS] correspond to those for the efficient influence function $\tau_1$ in Proposition 1. Proposition 3 collects theoretical comparisons between the three estimators.

PROPOSITION 3. *The following statements hold:*

(i) $\hat{\mu}_{AIPW,\text{fix}}$ is doubly robust.

(ii) $\hat{\mu}_{AIPW,\text{fix}}$ is locally efficient: if a PS and an OR model are correctly specified, then $\hat{\mu}_{AIPW,\text{fix}}$ achieves the semiparametric variance bound and hence

asy.var $(\hat{\mu}_{AIPW,\text{fix}}) \leq$ asy.var $(\hat{\mu}_{AIPW})$.

(iii) If an OR model is correctly specified and $m_1(X)$ is efficiently estimated in $\hat{\mu}_{OR}$, then

asy.var $(\hat{\mu}_{AIPW,\text{fix}}) \geq$ asy.var $(\hat{\mu}_{OR})$.

Compared with $\hat{\mu}_{OR}$, $\hat{\mu}_{AIPW,\text{fix}}$ is more robust in terms of bias if the OR model is misspecified but the PS model is correctly specified, but is less efficient in terms of variance if the OR model is correctly specified. The usual bias-variance trade-off takes effect. Compared with $\hat{\mu}_{AIPW}$, $\hat{\mu}_{AIPW,\text{fix}}$ is more robust in terms of bias if the PS model is misspecified but the OR model is correctly specified, and is more efficient in terms of variance if both the PS and the OR models are correctly specified. The usual bias-variance trade-off seems not to exist. Intuitively, the difference can be attributed to the characteristics of OR (being aggressive) and PS (being conservative) discussed in Section 1. It is possible for the PS approach to reduce both bias and variance by incorporating an OR model, but not so for the OR approach by incorporating a PS model.

Local efficiency implies that if the PS model is correctly specified, then $\hat{\mu}_{AIPW,\text{fix}}$ gains efficiency over $\hat{\mu}_{AIPW}$ for every function $h(X)$ under the condition that the OR model is also correctly specified. A more desirable situation is to find an estimator that is not only doubly robust and locally efficient but also, whenever the PS model is correctly specified, guaranteed to gain efficiency over $\hat{\mu}_{AIPW}$ for any initial, fixed function $h(X)$. For simplicity, consider $\hat{\mu}_{IPW}$ corresponding to $h(X) = 0$ as the initial estimator. In this case, consider Tan's (2006) regression (tilde) estimator

$$\tilde{\mu}_{REG} = \frac{1}{n}\sum_{i=1}^n \frac{T_i Y_i}{\hat{\pi}(X_i)} - \tilde{\beta}^{(1)} \frac{1}{n}\sum_{i=1}^n \left(\frac{T_i}{\hat{\pi}(X_i)} - 1\right) \hat{m}_1(X_i),$$

where $\tilde{\beta}^{(1)}$ is the first element of $\tilde{\beta} = \tilde{E}^{-1}(\hat{\xi}\hat{\zeta}^\top)\tilde{E}(\hat{\xi}\hat{\eta})$, $\tilde{E}$ denotes sample average, and

$$\hat{\eta} = \frac{TY}{\hat{\pi}(X)},$$

$$\hat{\xi} = \left(\frac{T}{\hat{\pi}(X)} - 1\right)\left(\hat{m}_1(X), \frac{\frac{\partial \hat{\pi}}{\partial \gamma^\top}(X)}{1 - \hat{\pi}(X)}\right)^\top,$$

$$\hat{\zeta} = \frac{T}{\hat{\pi}(X)}\left(\hat{m}_1(X), \frac{\frac{\partial \hat{\pi}}{\partial \gamma^\top}(X)}{1 - \hat{\pi}(X)}\right)^\top.$$

This estimator algebraically resembles Robins, Rotnitzky and Zhao's (1995) regression (hat) estimator

$$\hat{\mu}_{REG} = \frac{1}{n}\sum_{i=1}^n \frac{T_i Y_i}{\hat{\pi}(X_i)}$$
$$- \hat{\beta}^{(1)} \frac{1}{n}\sum_{i=1}^n \left(\frac{T_i}{\hat{\pi}(X_i)} - 1\right) \hat{m}_1(X_i),$$

where $\hat{\beta}^{(1)}$ is the first element of $\hat{\beta} = \tilde{E}^{-1}(\hat{\xi}\hat{\xi}^\top)\tilde{E}(\hat{\xi}\hat{\eta})$. Compared with $\hat{\mu}_{AIPW,\text{fix}}$, each estimator introduces an estimated regression coefficient, $\tilde{\beta}$ or $\hat{\beta}$, of $\hat{\eta}$ against control variates $\hat{\xi}$. Therefore, $\tilde{\mu}_{REG}$ and $\hat{\mu}_{REG}$ share the advantage of optimally using control variates $\hat{\xi}$ [Proposition 4(ii)]. See Section 3 for a discussion about "control variates" and "regression estimators." On the other hand, $\hat{\beta}$ is defined in the classical manner, whereas $\tilde{\beta}$ is specially constructed by exploiting the structure of control variates $\hat{\xi}$. This subtle difference underlies Proposition 4(i).

PROPOSITION 4. *The following statements hold:*

(i) $\tilde{\mu}_{REG}$ and $\hat{\mu}_{REG}$ are locally efficient, but $\tilde{\mu}_{REG}$ is doubly robust and $\hat{\mu}_{REG}$ is not.

(ii) If a PS model is correctly specified and $\pi(X)$ is efficiently estimated, then $\tilde{\mu}_{REG}$ and $\hat{\mu}_{REG}$ achieve the smallest asymptotic variance among

$$\frac{1}{n}\sum_{i=1}^n \frac{T_i Y_i}{\hat{\pi}(X_i)} - b^{(1)}\frac{1}{n}\sum_{i=1}^n \left(\frac{T_i}{\hat{\pi}(X_i)} - 1\right)\hat{m}_1(X_i),$$

where $b^{(1)}$ is an arbitrary coefficient. The two estimators are asymptotically at least as efficient as $\hat{\mu}_{IPW}$ and $\hat{\mu}_{AIPW,\text{fix}}$, corresponding to $b^{(1)} = 0$ and 1.

Compared with $\hat{\mu}_{AIPW,\text{fix}}$, $\tilde{\mu}_{REG}$ provides a more concrete improvement upon $\hat{\mu}_{IPW}$ due to the possession of three properties: optimality in using control variates, local efficiency and double robustness. Using $\tilde{\mu}_{REG}$ achieves variance reduction if the PS model is correctly specified (the effect of which is maximal if the OR model is also correctly specified), and bias reduction if the PS model is misspecified but the OR model is correctly specified. On the other hand, comparison between $\hat{\mu}_{OR}$ and $\tilde{\mu}_{REG}$ is similarly subject to the usual bias-variance trade-off as that between $\hat{\mu}_{OR}$ and $\tilde{\mu}_{AIPW,\text{fix}}$. That is, $\tilde{\mu}_{REG}$ is more robust than $\hat{\mu}_{OR}$ if the OR model is misspecified but the PS model is correctly specified, but is less efficient if the OR model is correctly specified.

The preceding comparisons between $\hat{\mu}_{AIPW,\text{fix}}$, $\tilde{\mu}_{REG}$ and $\hat{\mu}_{OR}$, $\hat{\mu}_{IPW}$ present useful facts for understanding DR estimation. It seems more meaningful to consider $\hat{\mu}_{AIPW,\text{fix}}$ or $\tilde{\mu}_{REG}$ as an advance or improvement in the PS approach by incorporating an OR model rather than in the OR approach by incorporating a PS model. The OR and PS models play different roles, even though the models are equally referred to in the concept of DR and $\hat{\mu}_{AIPW,\text{fix}}$ can be expressed as bias-corrected $\hat{\mu}_{OR}$ or equivalently as bias-corrected $\hat{\mu}_{IPW}$. This viewpoint is also supported by the construction of $\hat{\mu}_{AIPW,\text{fix}}$ (in the first expression by Robins, Rotnitzky and Zhao, 1994) and $\tilde{\mu}_{REG}$. Both of the estimators are derived under the assumption that the PS model is correct, and then examined in the situation where the OR model is also correct, or the PS model is misspecified but the OR model correct (see Tan, 2006, Section 3.2).

The different characteristics discussed in Section 1 persist between the PS (even using $\hat{\mu}_{AIPW,\text{fix}}$ or $\tilde{\mu}_{REG}$ with the DR benefit) and OR approaches. The asymptotic variance of $\hat{\mu}_{AIPW}$, $\hat{\mu}_{AIPW,\text{fix}}$, or $\tilde{\mu}_{REG}$ if $\tilde{\mu}_{REG}$ the PS model is correctly specified is no smaller, whereas that of $\hat{\mu}_{OR}$ if the OR model is correctly specified is no greater, than the semiparametric variance bound. Moreover, if the OR model is correct, the asymptotic variance of $\hat{\mu}_{AIPW,\text{fix}}$ or $\tilde{\mu}_{REG}$ is still no smaller than that of $\hat{\mu}_{OR}$. Therefore:

PROPOSITION 5. *The asymptotic variance of $\hat{\mu}_{AIPW,\text{fix}}$ or $\tilde{\mu}_{REG}$ if either a PS or an OR model is correctly specified is no smaller than that of $\hat{\mu}_{OR}$ if the OR model is correctly specified and $m_1(X)$ is efficiently estimated in $\hat{\mu}_{OR}$.*

Like Proposition 2, this result does *not* establish absolute superiority of the OR approach over the PS-DR approach. Instead, it points to considering practical issues of model specification and consequences of model misspecification. There seems to be no definite comparison, because various, unmeasurable factors are involved. Nevertheless, the points regarding questions (a) and (b) in Section 1 remain relevant.

In summary, it seems more constructive to view DR estimation in the PS approach by incorporating an OR model rather than in the OR approach by incorporating a PS model. The estimator $\tilde{\mu}_{REG}$ provides a concrete improvement upon $\hat{\mu}_{IPW}$ with both variance and bias reduction in the sense that it gains efficiency whenever the PS model is correctly specified (and maximally so if the OR model is also correctly specified), and remains consistent if the PS model is misspecified but the OR model is correctly specified. On the other hand, comparison between $\tilde{\mu}_{REG}$ and $\hat{\mu}_{OR}$ is complicated by the usual bias-variance trade-off. Different characteristics are associated with the OR and the PS-DR approaches and should be carefully weighed in applications.

## 3. OTHER COMMENTS

### Control Variates and Regression Estimators

The name "regression estimator" is adopted from the literatures of sampling survey (e.g., Cochran, 1977, Chapter 7) and Monte Carlo integration (e.g., Hammersley and Handscomb, 1964), and should be distinguished from "regression estimation" described by KS (Section 2.3). Specifically, the idea is to exploit the fact that if the PS model is correct, then $\hat{\eta}$ asymptotically has mean $\mu_1$ (to be estimated) and $\hat{\xi}$ mean 0 (known). That is, $\hat{\xi}$ serve as auxiliary variables (in the terminology of survey sampling) or control variates (in that of Monte Carlo integration). Variance reduction can be achieved by using $\tilde{E}(\hat{\eta}) - b\tilde{E}(\hat{\xi})$, instead of $\hat{\mu}_{IPW} = \tilde{E}(\hat{\eta})$, with $b$ an estimated regression coefficient of $\hat{\eta}$ against $\hat{\xi}$.

The control variates for $\tilde{\mu}_{REG}$ in Section 2 include $(\hat{\pi}^{-1}T - 1)\hat{m}_1$ and $(T - \hat{\pi})[\hat{\pi}(1 - \hat{\pi})]^{-1}\partial\hat{\pi}/\partial\gamma$, the second of which is the score function for the PS model and is necessary for asymptotic optimality in Proposition 4(ii). If the PS model is correct, then $\tilde{\mu}_{REG}$ is always at least as efficient as $\hat{\mu}_{IPW}$ in the raw version, that is, $\hat{\mu}_{AIPW}(0)$, but not always than $\hat{\mu}_{IPW}$ in the ratio version. However, the indefiniteness can be easily resolved. If the control variate $\hat{\pi}^{-1}T - 1$ is added, or $(1, \hat{m}_1)^\top$ substituted for $\hat{m}_1$, then $\tilde{\mu}_{REG}$ always gains efficiency over both versions of $\hat{\mu}_{IPW}$. Furthermore, if $(1, h, \hat{m}_1)^\top$ is substituted for $\hat{m}_1$, then $\tilde{\mu}_{REG}$ always gains efficiency also over the estimator $\hat{\mu}_{AIPW}(h)$.

### Causal Inference

Causal inference involves estimation of both $\mu_1$ and $\mu_0$. Similar estimators of $\mu_0$ can be separately defined by replacing $T$, $\hat{\pi}$ and $\hat{m}_1$ with $1 - T$, $1 - \hat{\pi}$ and $\hat{m}_0$, where $m_0 = E(Y|T=0,X)$. The control variates $((1-\hat{\pi})^{-1}(1-T)-1)(1,\hat{m}_0)^\top$ for estimating $\mu_0$ differ from $(\hat{\pi}^{-1}T-1)(1,\hat{m}_1)^\top$ for estimating $\mu_1$. As a consequence, even though $\tilde{\mu}_{1,REG}$ or $\tilde{\mu}_{0,REG}$ individually gains efficiency over $\hat{\mu}_{1,IPW}$ or $\hat{\mu}_{0,IPW}$, the difference $\tilde{\mu}_{1,REG} - \tilde{\mu}_{0,REG}$ does not necessarily gain efficiency over $\hat{\mu}_{1,IPW} - \hat{\mu}_{0,IPW}$. The problem can be overcome by using a combined set of control variates, say, $[\hat{\pi}^{-1}T - (1-\hat{\pi})^{-1}(1-T)](\hat{\pi}, 1-\hat{\pi}, \hat{\pi}\hat{m}_0, (1-\hat{\pi})\hat{m}_1)^\top$. Then $\tilde{\mu}_{1,REG} - \tilde{\mu}_{0,REG}$ maintains optimality in using control variates in the sense of Proposition 4(ii), in addition to local efficiency and double robustness. The mechanism of using a common set of control variates for estimating both $\mu_1$ and $\mu_0$ is automatic in the likelihood PS approach of Tan (2006).

## PS Stratification

KS (Section 2.2) described the stratification estimator of Rosenbaum and Rubin (1983) as a way "to coarsen the estimated propensity score into a few categories and compute weighted averages of the mean response across categories." It is helpful to rewrite the estimator in their display (6) as

$$\hat{\mu}_{\text{strat}} = \frac{1}{n}\sum_{i=1}^{n}\frac{T_i Y_i}{\hat{\pi}_{\text{strat}}(X_i)},$$

where $\hat{\pi}_{\text{strat}}(X) = \sum_{i=1}^{n} T_i 1\{\hat{\pi}(X_i) \in \hat{S}_j\}/\sum_{i=1}^{n} 1\{\hat{\pi}(X_i) \in \hat{S}_j\}$ if $\hat{\pi}(X) \in \hat{S}_j$ (the $j$th estimated PS stratum), $j=1,\ldots,s$. That is, $\hat{\mu}_{\text{strat}}$ is exactly an IPW estimator based on the discretized $\hat{\pi}_{\text{strat}}(X)$. Comparison between $\hat{\mu}_{\text{strat}}$ and $\hat{\mu}_{IPW}$ is subject to the usual bias-variance trade-off. On one hand, $\hat{\mu}_{\text{strat}}$ often has smaller variance than $\hat{\mu}_{IPW}$. On the other hand, the asymptotic limit of $\hat{\mu}_{\text{strat}}$ can be shown to be

$$\sum_{j=1}^{s} \frac{E[\pi(X)m_1(X)|\pi^*(X) \in S_j^*]}{E[\pi(X)|\pi^*(X) \in S_j^*]} P(\pi^*(X) \in S_j^*),$$

where $\pi^*(X)$ is the limit of $\hat{\pi}(X)$, which agrees with the true $\pi(X)$ if the PS model is correct, and $S_j^*$ is that of $\hat{S}_j$. The ratio inside the above sum is the within-stratum average of $m_1(X)$ weighted proportionally to $\pi(X)$. Therefore, $\hat{\mu}_{\text{strat}}$ is inconsistent unless $\pi(X)$ or $m_1(X)$ is constant within each stratum (cf. KS's discussion about crude DR in Section 2.4). The asymptotic bias depends on the joint behavior of $m_1(X)$ and $\pi(X)$, and can be substantial if $m_1(X)$ varies where $\pi(X) \approx 0$ varies so that $m_1(X)$ are weighted differentially, say, by a factor of 10 at two $X$'s with $\pi(X) = 0.01$ and $0.1$.

## Simulations

KS designed a simulation setup with an OR and a PS model appearing to be "nearly correct." The response is generated as $Y = 210 + 27.4Z_1 + 13.7Z_2 + 13.7Z_3 + 13.7Z_4 + \epsilon$, and the propensity score $\pi = \text{expit}(-Z_1 + 0.5Z_2 - 0.25Z_3 - 0.1Z_4)$, where $\epsilon$ and $(Z_1, Z_2, Z_3, Z_4)$ are independent, standard normal. The covariates seen by the statistician are $X_1 = \exp(Z_1/2)$, $X_2 = Z_2/(1+\exp(Z_1))+10$, $X_3 = (Z_1 Z_3/25 + 0.6)^3$ and $X_4 = (Z_2 + Z_4 + 20)^2$. The OR model is the linear model of $Y$ against $X$, and the PS model is the logistic model of $T$ against $X$.

In the course of replicating their simulations, we accidentally discovered that the following models also appear to be "nearly correct." The covariates seen by the statistician are the same $X_1, X_2, X_3$, but $X_4 = (Z_3 + Z_4 + 20)^2$. The OR model is linear and the PS model is logistic as KS models. For one simulated dataset, Figures 1 and 2 present scatterplots and boxplots similar to Figures 2 and 3 in KS. For the OR model, the regression coefficients are highly significant and $R^2 = 0.97$. The correlation between the fitted values of $Y$ under the correct and the misspecified OR models is 0.99, and that between the linear predictors under the correct and the misspecified PS models is 0.93. Tables 1 and 2 summarize our simulations for KS models and for the alternative models described above. The raw version of $\hat{\mu}_{IPW}$ is used. The estimators $\tilde{\mu}_{REG}^{(\text{m})}$ and $\hat{\mu}_{REG}^{(\text{m})}$ are defined as $\tilde{\mu}_{REG}$ and $\hat{\mu}_{REG}$ except that the score function for the PS model is dropped from $\hat{\xi}$. For these four estimators, $(1, \hat{m}_1)^\top$ is substituted for $\hat{m}_1$.

KS found that none of the DR estimators they tried improved upon the performance of the OR estimator; see also Table 1. This situation is consistent with the discussion in Section 2. The theory of DR estimation does *not* claim that a DR estimator is guaranteed to perform better than the OR estimator when the OR and the PS models are both misspecified, whether mildly or grossly. Therefore, KS's simulations serve as an example to remind us of this indefinite comparison.

On the other hand, neither is the OR estimator guaranteed to outperform DR estimators when the OR model is misspecified or even "nearly correct." As seen from Table 2, $\hat{\mu}_{OR}$ yields greater RMSE values than the DR estimators, $\hat{\mu}_{\text{WLS}}$, $\tilde{\mu}_{REG}$ and $\tilde{\mu}_{REG}^{(\text{m})}$ when the alternative, misspecified OR and PS mod-els are both used. For $n = 200$, the bias of $\hat{\mu}_{OR}$ is 2.5 and that of $\tilde{\mu}_{REG}$ is 0.44, which differ substantially from the corresponding biases $-0.56$ and $-1.8$ in Table 1 when KS models are used.

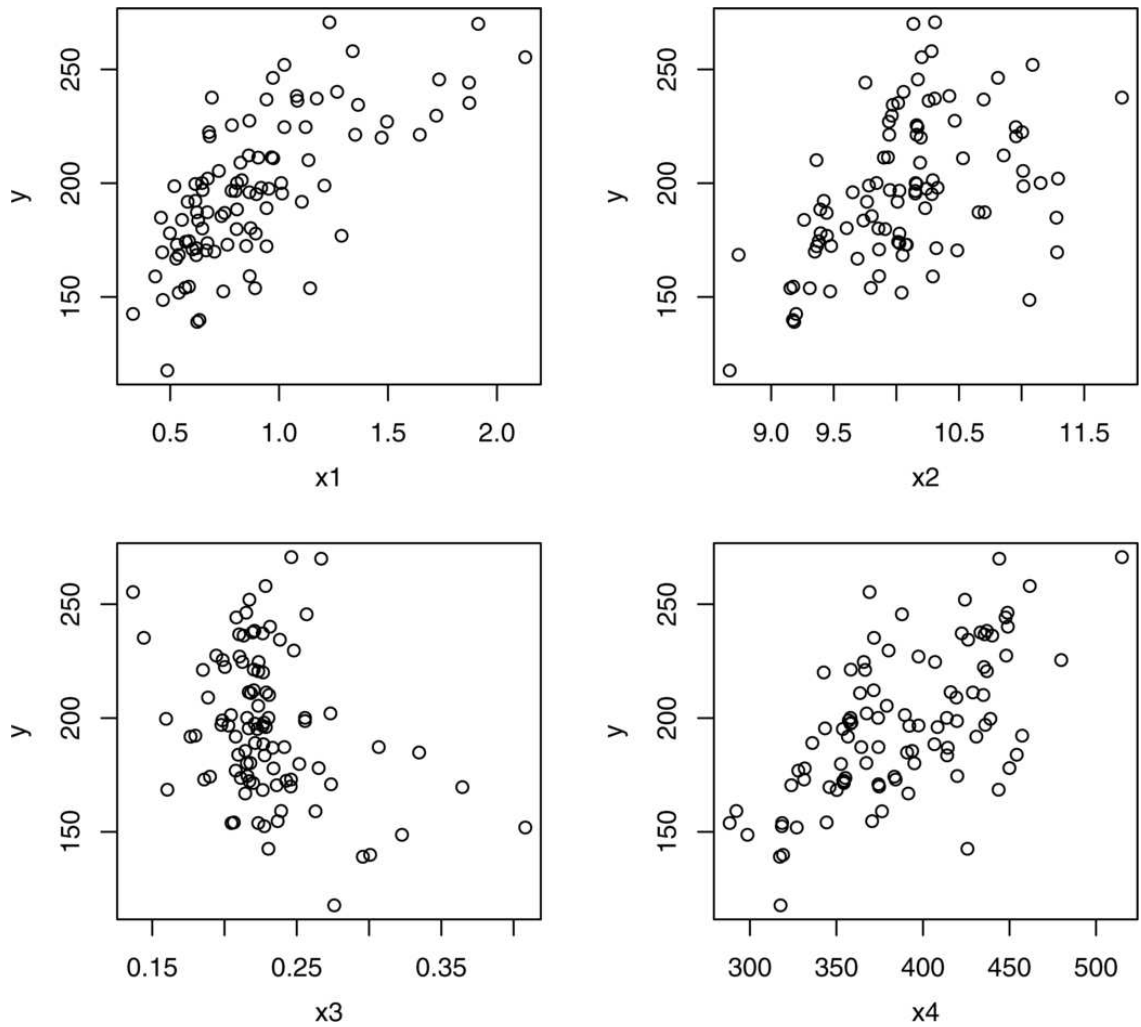

Fig. 1. *Scatterplots of response versus covariates (alternative models).*

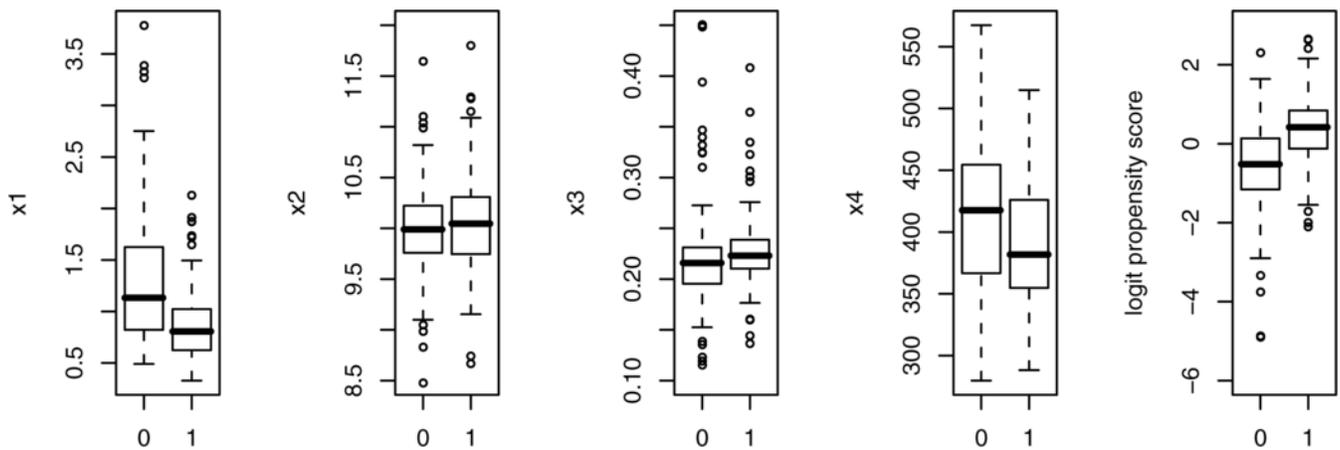

Fig. 2. *Boxplots of covariates and propensity scores (alternative models).*

TABLE 1
*Numerical comparison of estimators of $\mu_1$ (KS models)*

|  | Method | Bias | % Bias | RMSE | MAE | Bias | % Bias | RMSE | MAE |
|---|---|---|---|---|---|---|---|---|---|
| $n = 200$ |  | \multicolumn{4}{c}{$\pi$-model correct} |  |  | \multicolumn{2}{c}{$\pi$-model incorrect} |  |
|  | IPW | 0.080 | 0.64 | 12.6 | 6.11 | 16 | 32 | 52.7 | 8.99 |
|  | strat | $-1.1$ | $-37$ | 3.20 | 2.04 | $-2.9$ | $-93$ | 4.28 | 3.11 |
|  |  | \multicolumn{4}{c}{$y$-model correct} |  |  | \multicolumn{2}{c}{$y$-model incorrect} |  |
|  | OLS | $-0.025$ | $-0.99$ | 2.47 | 1.68 | $-0.56$ | 17 | 3.33 | 2.19 |
|  |  | \multicolumn{4}{c}{$y$-model correct} |  |  | \multicolumn{2}{c}{$y$-model incorrect} |  |
| $\pi$-model | AIPW$_{\text{fix}}$ | $-0.024$ | $-0.96$ | 2.47 | 1.67 | 0.24 | 6.9 | 3.44 | 2.06 |
| correct | WLS | $-0.025$ | $-1.0$ | 2.47 | 1.68 | 0.39 | 13 | 2.99 | 1.89 |
|  | REG$_{\text{tilde}}$ | $-0.025$ | $-1.0$ | 2.47 | 1.69 | 0.14 | 5.2 | 2.73 | 1.76 |
|  | REG$_{\text{hat}}$ | $-0.52$ | $-20$ | 2.63 | 1.73 | $-0.52$ | $-19$ | 2.81 | 1.78 |
|  | REG$_{\text{tilde}}^{(m)}$ | $-0.024$ | $-0.98$ | 2.47 | 1.68 | 0.24 | 8.9 | 2.74 | 1.79 |
|  | REG$_{\text{hat}}^{(m)}$ | $-0.21$ | $-8.4$ | 2.48 | 1.68 | $-0.086$ | $-3.2$ | 2.65 | 1.74 |
| $\pi$-model | AIPW$_{\text{fix}}$ | $-0.026$ | $-1.0$ | 2.48 | 1.71 | $-5.1$ | $-44$ | 12.6 | 3.75 |
| incorrect | WLS | $-0.026$ | $-1.0$ | 2.47 | 1.70 | $-2.2$ | $-69$ | 3.91 | 2.77 |
|  | REG$_{\text{tilde}}$ | $-0.027$ | $-1.1$ | 2.47 | 1.71 | $-1.8$ | $-62$ | 3.47 | 2.41 |
|  | REG$_{\text{hat}}$ | $-0.45$ | $-18$ | 2.60 | 1.71 | $-2.2$ | $-76$ | 3.68 | 2.53 |
|  | REG$_{\text{tilde}}^{(m)}$ | $-0.026$ | $-1.1$ | 2.47 | 1.69 | $-2.0$ | $-68$ | 3.56 | 2.47 |
|  | REG$_{\text{hat}}^{(m)}$ | $-0.13$ | $-5.3$ | 2.48 | 1.68 | $-2.2$ | $-77$ | 3.68 | 2.59 |
| $n = 1000$ |  | \multicolumn{4}{c}{$\pi$-model correct} |  |  | \multicolumn{2}{c}{$\pi$-model incorrect} |  |
|  | IPW | 0.098 | 2.0 | 4.98 | 3.04 | 68 | 9.2 | 746 | 14.7 |
|  | strat | $-1.1$ | $-86$ | 1.71 | 1.24 | $-2.9$ | $-214$ | 3.22 | 2.94 |
|  |  | \multicolumn{4}{c}{$y$-model correct} |  |  | \multicolumn{2}{c}{$y$-model incorrect} |  |
|  | OLS | $-0.047$ | $-4.0$ | 1.15 | 0.770 | $-0.85$ | $-56$ | 1.75 | 1.15 |
|  |  | \multicolumn{4}{c}{$y$-model correct} |  |  | \multicolumn{2}{c}{$y$-model incorrect} |  |
| $\pi$-model | AIPW$_{\text{fix}}$ | $-0.046$ | $-4.0$ | 1.15 | 0.766 | 0.043 | 2.6 | 1.63 | 1.11 |
| correct | WLS | $-0.046$ | $-4.0$ | 1.15 | 0.769 | 0.12 | 8.7 | 1.37 | 0.943 |
|  | REG$_{\text{tilde}}$ | $-0.046$ | $-4.0$ | 1.15 | 0.773 | 0.048 | 3.9 | 1.23 | 0.809 |
|  | REG$_{\text{hat}}$ | $-0.13$ | $-11$ | 1.16 | 0.796 | $-0.077$ | $-6.3$ | 1.23 | 0.812 |
|  | REG$_{\text{tilde}}^{(m)}$ | $-0.046$ | $-4.0$ | 1.15 | 0.770 | 0.092 | 7.3 | 1.26 | 0.870 |
|  | REG$_{\text{hat}}^{(m)}$ | $-0.083$ | $-7.2$ | 1.15 | 0.768 | 0.024 | 1.9 | 1.24 | 0.857 |
| $\pi$-model | AIPW$_{\text{fix}}$ | $-0.10$ | $-6.5$ | 1.61 | 0.769 | $-26$ | $-8.5$ | 308 | 5.56 |
| incorrect | WLS | $-0.048$ | $-4.1$ | 1.15 | 0.764 | $-3.0$ | $-203$ | 3.38 | 3.05 |
|  | REG$_{\text{tilde}}$ | $-0.046$ | $-4.0$ | 1.15 | 0.764 | $-1.7$ | $-120$ | 2.21 | 1.73 |
|  | REG$_{\text{hat}}$ | $-0.045$ | $-3.9$ | 1.16 | 0.786 | $-1.7$ | $-122$ | 2.24 | 1.75 |
|  | REG$_{\text{tilde}}^{(m)}$ | $-0.046$ | $-4.0$ | 1.15 | 0.763 | $-2.1$ | $-152$ | 2.48 | 2.04 |
|  | REG$_{\text{hat}}^{(m)}$ | $-0.058$ | $-5.0$ | 1.16 | 0.771 | $-2.2$ | $-158$ | 2.57 | 2.15 |

The consequences of model misspecification are difficult to study, because the degree and direction of model misspecification are subtle, even elusive. For the dataset examined earlier, the absolute differences between the (highly correlated) fitted values of $Y$ under the correct and the alternative, misspecified OR models present a more serious picture of model misspecification. In fact, the quartiles of these absolute differences are 2.0, 3.2 and 5.1, and the maximum is 20.

For both Tables 1 and 2, the DR estimators $\tilde{\mu}_{REG}$ and $\tilde{\mu}_{REG}^{(m)}$ perform overall better than the other DR estimators $\hat{\mu}_{AIPW,\text{fix}}$ and $\hat{\mu}_{WLS}$. Compared with $\hat{\mu}_{WLS}$, $\tilde{\mu}_{REG}$ has MSE reduced by 15–20% (Table 1) or by 20–25% (Table 2) when the PS model is correct but the OR model is misspecified, which agrees with the optimality property of $\tilde{\mu}_{REG}$ in Proposition 4(ii). Even the simplified estimator $\tilde{\mu}_{REG}^{(m)}$ gains similar efficiency, although the gain is not guaranteed in theory. The non-DR estimators $\hat{\mu}_{REG}$ and $\hat{\mu}_{REG}^{(m)}$ sometimes have sizeable biases even when the PS model is correct.

TABLE 2
*Numerical comparison of estimators of $\mu_1$ (alternative models)*

|  | Method | Bias | % Bias | RMSE | MAE | Bias | % Bias | RMSE | MAE |
|---|---|---|---|---|---|---|---|---|---|
| $n=200$ | | | $\pi$-model correct | | | | $\pi$-model incorrect | | |
| | IPW | 0.080 | 0.64 | 12.6 | 6.11 | 18 | 34 | 55.7 | 9.61 |
| | strat | $-1.1$ | $-37$ | 3.20 | 2.04 | $-1.1$ | $-36$ | 3.22 | 2.21 |
| | | | $y$-model correct | | | | $y$-model incorrect | | |
| | OLS | $-0.025$ | $-0.99$ | 2.47 | 1.68 | 2.5 | 80 | 4.04 | 2.73 |
| | | | $y$-model correct | | | | $y$-model incorrect | | |
| $\pi$-model | AIPW$_{\text{fix}}$ | $-0.024$ | $-0.96$ | 2.47 | 1.67 | 0.53 | 14 | 3.82 | 2.32 |
| correct | WLS | $-0.025$ | $-1.0$ | 2.47 | 1.68 | 0.83 | 28 | 3.09 | 1.96 |
| | REG$_{\text{tilde}}$ | $-0.025$ | $-1.0$ | 2.47 | 1.69 | 0.33 | 13 | 2.63 | 1.71 |
| | REG$_{\text{hat}}$ | $-0.52$ | $-20$ | 2.63 | 1.73 | $-0.34$ | $-13$ | 2.70 | 1.74 |
| | REG$_{\text{tilde}}^{(m)}$ | $-0.024$ | $-0.98$ | 2.47 | 1.68 | 0.45 | 17 | 2.74 | 1.78 |
| | REG$_{\text{hat}}^{(m)}$ | $-0.21$ | $-8.4$ | 2.48 | 1.68 | 0.09 | 3.6 | 2.63 | 1.74 |
| $\pi$-model | AIPW$_{\text{fix}}$ | $-0.024$ | $-0.97$ | 2.48 | 1.71 | $-2.5$ | $-21$ | 12.2 | 2.72 |
| incorrect | WLS | $-0.026$ | $-0.10$ | 2.47 | 1.70 | 0.33 | 11 | 3.11 | 2.05 |
| | REG$_{\text{tilde}}$ | $-0.025$ | $-0.10$ | 2.47 | 1.71 | 0.44 | 16 | 2.74 | 1.80 |
| | REG$_{\text{hat}}$ | $-0.42$ | $-17$ | 2.56 | 1.71 | $-0.026$ | $-0.95$ | 2.74 | 1.78 |
| | REG$_{\text{tilde}}^{(m)}$ | $-0.025$ | $-1.0$ | 2.47 | 1.69 | 0.31 | 11 | 2.83 | 1.80 |
| | REG$_{\text{hat}}^{(m)}$ | $-0.22$ | $-8.9$ | 2.48 | 1.71 | 0.035 | 1.3 | 2.76 | 1.77 |
| $n=1000$ | | | $\pi$-model correct | | | | $\pi$-model incorrect | | |
| | IPW | 0.098 | 2.0 | 4.98 | 3.04 | 80 | 8.5 | 951 | 16.8 |
| | strat | $-1.1$ | $-86$ | 1.71 | 1.24 | $-0.96$ | $-72$ | 1.65 | 1.17 |
| | | | $y$-model correct | | | | $y$-model incorrect | | |
| | OLS | $-0.047$ | $-4.0$ | 1.15 | 0.770 | 2.2 | 152 | 2.67 | 2.21 |
| | | | $y$-model correct | | | | $y$-model incorrect | | |
| $\pi$-model | AIPW$_{\text{fix}}$ | $-0.046$ | $-4.0$ | 1.15 | 0.766 | 0.061 | 3.3 | 1.87 | 1.17 |
| correct | WLS | $-0.046$ | $-4.0$ | 1.15 | 0.769 | 0.22 | 16 | 1.39 | 0.957 |
| | REG$_{\text{tilde}}$ | $-0.046$ | $-4.0$ | 1.15 | 0.773 | 0.12 | 10 | 1.21 | 0.818 |
| | REG$_{\text{hat}}$ | $-0.13$ | $-11$ | 1.16 | 0.796 | $-0.012$ | $-0.97$ | 1.19 | 0.801 |
| | REG$_{\text{tilde}}^{(m)}$ | $-0.046$ | $-4.0$ | 1.15 | 0.770 | 0.14 | 12 | 1.25 | 0.849 |
| | REG$_{\text{hat}}^{(m)}$ | $-0.083$ | $-7.2$ | 1.15 | 0.768 | 0.069 | 5.7 | 1.22 | 0.826 |
| $\pi$-model | AIPW$_{\text{fix}}$ | $-0.12$ | $-6.3$ | 1.83 | 0.780 | $-31$ | $-6.9$ | 441 | 2.92 |
| incorrect | WLS | $-0.048$ | $-4.1$ | 1.15 | 0.768 | $-0.55$ | $-38$ | 1.55 | 1.12 |
| | REG$_{\text{tilde}}$ | $-0.044$ | $-3.9$ | 1.15 | 0.765 | 0.61 | 46 | 1.46 | 0.946 |
| | REG$_{\text{hat}}$ | $-0.099$ | $-8.5$ | 1.16 | 0.787 | 0.57 | 43 | 1.45 | 0.910 |
| | REG$_{\text{tilde}}^{(m)}$ | $-0.045$ | $-3.9$ | 1.15 | 0.757 | 0.22 | 17 | 1.29 | 0.847 |
| | REG$_{\text{hat}}^{(m)}$ | $-0.16$ | $-14$ | 1.17 | 0.764 | 0.13 | 10 | 1.28 | 0.836 |

## Summary

One of the main points of KS is that two (moderately) misspecified models are not necessarily better than one. This point is valuable. But at the same time, neither are two misspecified models necessarily worse than one. Practitioners may choose to implement either of the OR and the PS-DR approaches, each with its own characteristics. It is helpful for statisticians to promote a common, rigorous understanding of each approach and to investigate new ways for improvement. We welcome KS's article and the discussion as a step forward in this direction.

## ACKNOWLEDGMENTS

We thank Xiao-Li Meng and Dylan Small for helpful comments.